\documentclass{article}
\usepackage{ltwol2e}
\usepackage{epsfig}
\usepackage{axodraw}

\def\gsim{\ \rlap{\raise 3pt \hbox{$>$}}{\lower 3pt \hbox{$\sim$}}\ }
\def\lsim{\ \rlap{\raise 3pt \hbox{$<$}}{\lower 3pt \hbox{$\sim$}}\ }
\newcommand{\bfig}{\begin{center}\begin{picture}}
\newcommand{\efig}[1]{\end{picture}\\{\small #1}\end{center}}
\newcommand{\flin}[2]{\ArrowLine(#1)(#2)}
\newcommand{\wlin}[2]{\DashLine(#1)(#2){3}}

\newcommand{\glin}[3]{\Photon(#1)(#2){2}{#3}}

\newcommand{\sof}{\SetOffset}
\newcommand{\sca}{\SetScale}

\def\be{\begin{equation}}
\def\ee{\end{equation}}
\def\nn{\nonumber}
\def\bea{\begin{eqnarray}}
\def\eea{\end{eqnarray}}

\begin{document}

\begin{titlepage}

\begin{flushright}
CERN-TH/98-305\\LPTHE Orsay-98/59\\
hep-ph/yymmddd
\end{flushright}

\vspace{2.5cm}

\begin{center}
\Large\bf 
Five-leg photon--neutrino interactions
\end{center}

\vspace{1.2cm}

\begin{center}
A. Abada, 
J. Matias  and R. Pittau\\
{\sl Theory Division, CERN, CH-1211 Geneva 23, Switzerland}
\end{center}

\vspace{1.3cm}

\begin{center}
{\bf Abstract:}\\[0.3cm]
\parbox{11cm}{In a first part, we justify the feasibility 
of substituting a photon leg by a neutrino current in the Euler--Heisenberg
 Lagrangian to obtain an effective Lagrangian for the process 
  $\gamma\nu\to\gamma\gamma \nu$ and its crossed reactions. 
  We establish the link between these processes and the four-photon scattering 
  in both the Standard Model and the effective theory. As an application, 
  we compute in this effective theory the processes $\gamma\nu\to\gamma\gamma\nu$ and  
$\gamma\gamma\to\gamma{\nu}\bar\nu$ and show how to use the $\gamma\gamma\to\gamma\gamma$ 
results as a check. We settle the question of  the disagreement between 
two computations in the literature concerning the reaction
$\gamma\gamma\to \gamma \nu\overline\nu$.
In the second part, we present results of the direct computation of the
photon--neutrino five-leg processes in the Standard Model,  discuss 
possible astrophysical implications of our results, and
provide simple fits to the exact expressions.
}
\end{center}

\vspace{1cm}

\begin{center}
{\sl Contributed paper to\\
XXIXth International Conference on High Energy Physics\\
Vancouver, B.C., Canada, 23--29 July 1998}\\{abstract 1025}
\end{center}

\vfil
\noindent
CERN-TH/98-305\\
September 1998

\end{titlepage}

\setcounter{page}{1}

\title{FIVE-LEG PHOTON--NEUTRINO INTERACTIONS}

\author{A. Abada, J. Matias, R. Pittau}

\address{Theory Division, CERN, CH-1211 Geneva 23, Switzerland\\E-mail: 
abada@mail.cern.ch, matias@mail.cern.ch, pittau@mail.cern.ch}


\twocolumn[\maketitle\abstracts{In a first part, we justify the feasibility 
of substituting a photon leg by a neutrino current in the Euler--Heisenberg
 Lagrangian to obtain an effective Lagrangian for the process 
  $\gamma\nu\to\gamma\gamma \nu$ and its crossed reactions. 
  We establish the link between these processes and the four-photon scattering 
  in both the Standard Model and the effective theory. As an application, 
  we compute in this effective theory the processes $\gamma\nu\to\gamma\gamma\nu$ and  
$\gamma\gamma\to\gamma{\nu}\bar\nu$ and show how to use the $\gamma\gamma\to\gamma\gamma$ 
results as a check. We settle the question of  the disagreement between 
two computations in the literature concerning the reaction
$\gamma\gamma\to \gamma \nu\overline\nu$.
In the second part, we present results of the direct computation of the
photon--neutrino five-leg processes in the Standard Model,  discuss 
possible astrophysical implications of our results, and
provide simple fits to the exact expressions.}]

\section{Introduction}
Processes involving photons and neutrinos are potentially of interest in 
astrophysics and cosmology. 
However, it was realized some time ago in \cite{Liu} that 4-leg processes
($\gamma\nu   \to \gamma\nu$,
 $\gamma\gamma    \to \nu\bar\nu$ and  $ \nu\bar\nu   \to \gamma\gamma$) 
are too strongly suppressed to be of relevance. The reason for this 
suppression is the prohibition of a  two-photon coupling to a $J=1$
state. This is because of Yang's theorem \cite{Yang}.
On the other hand, this theorem does not apply to 5-leg processes
involving two neutrinos 
and three photons, such as
\begin{eqnarray}
\gamma\nu    &\to& \gamma\gamma\nu     \nonumber \\
\gamma\gamma &\to& \gamma\nu\bar\nu    \nonumber \\
\nu\bar\nu   &\to& \gamma\gamma\gamma  \label{eq4}\,, 
\end{eqnarray} 
The extra $\alpha$ in the cross section is 
compensated by an interchange of the  
 $1/M_W$ suppression
by an 
$1/m_e$ enhancement \cite{dic3,nous}. The relative enhancement of the
5-leg process versus the 4-leg one happens to be of several orders of
magnitude, depending on the energy.

In  \cite{dic3}, Dicus and Repko derived an effective Lagrangian for the 
above five-leg photon--neutrino interactions  
by using  the Euler--Heisenberg
Lagrangian that describes the photon--photon scattering \cite{EH}.
Moreover, in the literature there already existed  a computation
by Shabalin and Hieu \cite{Shabalin}
 of the second 
process in  (\ref{eq4}) whose result disagrees with the one given
in \cite{dic3}.

To settle this question, in a recent work \cite{nous} we computed
the first and the second processes \mbox{ (\ref{eq4})}, 
in the framework of the effective theory, 
confirming the results reported in ref. \cite{dic3}. 
In section 2, we justify the feasibility of this approach \cite{nous}
and give another derivation of the 5-leg effective vertex starting 
from the Euler--Heisenberg Lagrangian \cite{EH}.

The effective approach gives reliable results for energies below 
the threshold for $e^+ e^-$ pair production, while its
extrapolation to energies above $1$ MeV, interesting to study, for example, 
supernova dynamics, is suspect. Therefore, an exact calculation 
of the processes (\ref{eq4}) is 
important to see their role in astrophysics
and the range of validity of the effective theory.
This calculation \cite{nous2} is summarized in section 3, with the main results.

Very recently, parallel work in the same direction has been carried out 
by Dicus, Kao and Repko \cite{dic1}, so that we had a chance to compare our numerical 
results, finding complete agreement between the two independent calculations.

We end up with a discussion on some of the implications of the exact
results of these 5-leg processes 
in astrophysics and cosmology, and we give our conclusions.

\section{Effective theory}

The starting point is the leading term of the Euler--Heisenberg 
Lagrangian \cite{EH}, which describes the photon--photon scattering of 
Fig. 1a:
\bea 
{\cal L}_{{\mathrm {E-H}}}={\alpha^2\over 180 m_e^4}
\left [ 5\left(F_{\mu\nu}F^{\mu\nu}\right)^2 
-14 F_{\mu\nu}F^{\nu\lambda}F_{\lambda\rho}F^{\rho\mu}\right ]
\label{e-h} 
\eea
where $F_{\mu\nu}$ is the  photon field-strength tensor and $\alpha$ the QED
coupling constant.\\

\sca{0.8}
\bfig(300,80)
\sof(-20,20)
\GOval(110,40)(4,4)(0){0}
\glin{110,0}{110,40}{5}
\glin{110,40}{65,40}{5}
\glin{110,40}{110,85}{5}
\glin{110,40}{155,40}{5}  
\Text(92,10)[l]{$\gamma$}   
\Text(92,56)[l]{$\gamma$}
\Text(61,28)[t]{$\gamma$} 
\Text(114,28)[t]{$\gamma$}
\Text(60,40)[r]{$p_1,~\epsilon_1$} 
\Text(114,40)[l]{$p_3,~\epsilon_3$}
\Text(88,72)[b]{$p_2,~\epsilon_2$}
\Text(88,-2)[t]{$p_4,~\epsilon_4$}
\Text(130,67)[bl]{$(a)$}
\sof(100,20)
\glin{110,0}{110,20}{3}
\flin{110,20}{90,40}
\flin{90,40}{110,60}
\flin{110,60}{130,40}
\flin{130,40}{110,20}
\glin{90,40}{65,40}{3}
\glin{110,60}{110,85}{3}
\glin{130,40}{155,40}{3}  
\Text(92,10)[l]{$\gamma$}   
\Text(106,46)[r]{$e$}
\Text(114,24)[l]{$\gamma$}
\Text(61,28)[t]{$\gamma$} 
\Text(92,62)[t]{$\gamma$}
\Text(60,40)[r]{$p_1,~\epsilon_1$} 
\Text(114,40)[l]{$p_3,~\epsilon_3$}
\Text(88,72)[b]{$p_2,~\epsilon_2$}
\Text(88,-2)[t]{$p_4,~\epsilon_4$}
\Text(130,67)[bl]{$(b)$}
\efig{Figure. 1:  Four-photon interaction (a) in the effective Lagrangian 
and (b) in the SM.}
In what follows we will show the
relation between the effective Lagrangian of eq.
 (\ref{e-h}) and the one describing processes
(\ref{eq4}), which is given by 
\bea \label{our}
\!\!{\cal L}_{\rm eff}\!=\!{C \over 180}\!\!
\left [ \!5\left({\tilde F}_{\mu\nu}F^{\mu\nu}   \!\!  \right)\!
          \left(\!F _{\lambda\rho} F^{\lambda\rho}\!\right)
-14\! {\tilde F}_{\mu\nu}F^{\nu\lambda}F_{\lambda\rho}F^{\rho\mu}\!\right ]
\eea
where 
\bea
C={g^5 s_{W}^{3} (1 + v_{e}) \over 32 \pi^{2} m_{e}^{4} M_{W}^{2}}=\ {2 G_F
\alpha^{3/2}(1 + v_{e}) \over \sqrt{2\pi} m_e^4}\label{prefactor}\,,
\eea
and $\tilde F_{\mu \nu}$ stands for the field strength of the new 
``gauge field'' ${\tilde A}_{\nu}\equiv{\bar \psi} \gamma_{\nu}(1 -
\gamma_{5}) \psi=2 \Gamma_{\nu}$, with $\Gamma_{\nu}$ the neutrino current
$$\Gamma_{\mu}={\bar v}_{+}(5) \gamma_{\mu} w_{-} u_{-}(4)\ .$$
We use the notation $w^{\pm}=(1 \pm \gamma_{5})/2$ and  
$v_{e}=- 1/ 2+2 s_{W}^{2}$, where $s_W$ is the sine of the Weinberg angle.

\vskip 0.2cm

\sca{0.8}
\bfig(300,80)
\sof(-40,10)
\flin{40,0}{110,0}  \flin{110,0}{180,0}
\wlin{110,0}{110,20}\
\flin{110,20}{90,40}
\flin{90,40}{110,60}
\flin{110,60}{130,40}
\flin{130,40}{110,20}
\glin{90,40}{65,40}{3}
\glin{110,60}{110,85}{3}
\glin{130,40}{155,40}{3}
\Text(63,-5)[t]{$p_4~~~~\nu$}
\Text(124,-5)[t]{$\nu~~~~p_5$}
\Text(92,12)[l]{$Z$}
\Text(92,56)[l]{$\gamma$}
\Text(61,29)[t]{$\gamma$}
\Text(114,29)[t]{$\gamma$}
\Text(48,32)[r] {$\{i\}$}
\Text(128,32)[l]{$\{k\}$}
\Text(88,72)[b]{$\{j\}$}
\Text(104,44)[r]{$e$}
\Text(130,67)[bl]{$(a)$}
\sof(100,10)
\flin{40,0}{80,0}  \wlin{80,0}{140,0} \flin{140,0}{180,0}
\flin{80,0}{90,20} \flin{90,20}{110,32}
\flin{110,32}{130,20} \flin{130,20}{140,0}
\glin{90,20}{67,37}{3}
\glin{110,32}{110,62}{3}
\glin{130,20}{153,37}{3}
\Text(51,-5)[t]{$p_4~~~~\nu$}
\Text(124,-5)[t]{$\nu~~~~p_5$}
\Text(90,4)[b]{$W$}
\Text(94,40)[l]{$\gamma$}
\Text(55,23)[t]{$\gamma$} 
\Text(120,23)[t]{$\gamma$}
\Text(52,36)[r] {$\{i\}$}
\Text(124,36)[l]{$\{k\}$}
\Text(88,56)[b]{$\{j\}$}
\Text(105,26)[r]{$e$}
\Text(130,67)[bl]{$(b)$}
\efig{Figure 2. SM leading diagrams contributing to five-leg 
photon--neutrino processes.}

The keypoint of the proof is to demonstrate that 
the SM amplitudes of  
diagrams (a) and (b) of Fig. 2, which we called $A_{ijk}$ and $B_{ijk}$,
respectively, can be rewritten in terms of diagram (b) of Fig. 1.

The total amplitude $M(\lambda,\rho,\sigma)$ of the 5-leg process reads
\bea\label{M}
M(\lambda, \rho, \sigma) 
&=& \left [(A_{123}+ A_{321})+
(B_{123}+ B_{321}) \right.
\nonumber \\
&+& 
(A_{132}+ A_{231})+
(B_{132}+ B_{231})+
\nonumber \\
&+& \left.
(A_{213}+ A_{312})+
(B_{213}+ B_{312})
\right ]  
\, , \eea
where we denote the photon polarization by $\lambda$,
$\rho$ and  $\sigma$ (see \cite{nous} for the explicit form of $A_{ijk}$
and $B_{ijk}$).

 The reason why the terms in \mbox{eq. (\ref{M})} are collected in pairs
is because, when adding the two terms in each pair, 
and using the reversing invariance of the $\gamma$-matrix traces,
changing $q$ to $-q$ and $m_e$ to $-m_e$, 
the $\gamma_5$ piece of the $Zee$ vertex cancels \cite{nous}. 
For instance, the couple
\bea
A_{123}+A_{321}=\epsilon^{\alpha}(\vec{P}_{1},\lambda)
\epsilon^{\beta}(\vec{P}_{2},\rho)
\epsilon^{\gamma}(\vec{P}_{3},\sigma)( A_{123}^{\alpha \beta \gamma}
+A_{321}^{\alpha \beta \gamma}) ,\nonumber
\eea
where
\bea \label{as}
A_{123}^{\alpha \beta \gamma}+A_{321}^{\alpha \beta \gamma}=-{2 \over {(2
\pi)}^4}
{(g
s_{W})}^3 {\left( {g \over 2 c_{W}}
\right)}^2
v_{e}\Gamma_{\mu} {1 \over \Delta_{Z}} L_{1}^{\mu \alpha
\beta \gamma} \nonumber \nonumber\eea
and 
\bea
L_{1}^{\mu \alpha \beta \gamma}=  \int dq^{n}
{\rm \,Tr\,}\left[ \gamma^{\mu} {1
\over
Q_{23}^{-}} \gamma^{\gamma} {1 \over Q_{2}^{-}} \gamma^{\beta} {1 \over
Q_{0}^{-}}
\gamma^{\alpha} {1 \over Q_{1}^{-}} \right],\nonumber
\eea cancels completely its axial part. We are using the notations 
of Refs. \cite{nous,nous2}:
\bea
& Q_{\pm i}^{\mp}= \rlap/Q_{\pm i} \mp m_{e}\,, &~~~Q_{\pm (ij)}^{\mp}= \rlap/Q_{\pm (ij)} \mp m_{e}\,, \nonumber \\
& Q_{\pm i}= Q_0 \pm p_i\,,~~~                  &~~~Q_{\pm (ij)}= Q_0 \pm p_i \pm p_j\,,~~~Q_0= q\,, \nonumber \\
& D_{\pm i}= Q_{\pm i}^+\cdot Q_{\pm i}^-\,,    &~~~D_{\pm (ij)}= Q_{\pm (ij)}^+\cdot Q_{\pm (ij)}^-\,.
\eea
and $1/ \Delta_{Z}=1/\left({(p_{4}+p_{5})}^{2}-M_{Z}^{2}
\right )\sim-1/M_{Z}^{2}$. 
 
The same trick can be applied to the couples of $B$'s. Let us take, 
for instance,
\bea \label{bs}
&&B_{123}^{\alpha \beta \gamma}+B_{321}^{\alpha \beta \gamma}=-{4
\over {(2 \pi)}^4}
{(g
s_{W})}^3 {\left( {g \over 2 \sqrt{2}}
\right)}^2
 \Gamma_{\mu} 
 L_{2}^{\mu \alpha \beta \gamma} \\
&&L_{2}^{\mu \alpha \beta \gamma}\!= \!\!\int \!\! dq^{n}
{\rm \,Tr\,}\left[ \gamma^{\mu} {1
\over
Q_{23}^{-}} \gamma^{\gamma} {1 \over Q_{2}^{-}} \gamma^{\beta} {1 \over
Q_{0}^{-}}
\gamma^{\alpha} {1 \over Q_{1}^{-}} \right]{1 \over \Delta_{W}(q)}\,\nonumber
\eea
where $\Delta_{W}(q)=(q+p_{2}+p_{3}+p_{5})^{2}-M_{W}^{2}$.

The second step of the proof is to shrink diagram (b) of Fig. 2 to diagram
(b) of Fig. 1 with the photon leg substituted by the neutrino current.
This is done by expanding the $W$ propagator inside the loop, using
\bea \label{split}
{1 \over \Delta_{W}(q)}={1 \over {q}^{2}-M_{W}^{2}} -
{k^{2} + 2 q \cdot k \over (q^{2}-M_{W}^{2}) ({(q+k)}^{2}
 -M_{W}^{2})}
\eea
where $\Delta_{W}(q)\equiv (q+k)^{2}-M_{W}^{2}$ and 
$k=p_{2}+p_{3}+p_{5}$. Once introduced in eq. (\ref{bs}), 
the first term in the r.h.s. of eq. (\ref{split}) 
gives rise to an $L_1$ type of integral plus a contribution
of next order in $1/M_{W}^2$. 
On the other hand, the second term on the r.h.s. of eq. (\ref{split}) 
is of order $1/M_{W}^{4}$, and can therefore be neglected.

In conclusion, at  leading order in $1/M_{W}^{2}$, the  set 
of four diagrams (from a total of 12) is
always proportional to $L_{1}$:
\bea \label{set1}
A_{123}^{\alpha \beta \gamma}+A_{321}^{\alpha \beta \gamma}+
B_{123}^{\alpha \beta \gamma}+B_{321}^{\alpha \beta \gamma}=
-{g^{5} s_{W}^3 \over 2} {(1+v_{e})
\over \Delta_{Z} c_W^{2}}
 \Gamma_{\mu} L_{1}^{ \mu \alpha \beta \gamma} \nonumber
\eea

We have shown that in the SM the 5-leg processes (\ref{eq4})
at leading order correspond up to a global factor to the 4-photon
scattering process; both theories should thus be described at this order
by the same effective theory, after substitution of  one polarization
vector by the neutrino current
\bea
\epsilon_{\mu}({\overrightarrow{P_4}},\lambda_{4})  
\rightarrow {1 \over 2} \Gamma_{\mu} \eea
and fixing the overall constant $C$. This constant $C$ (eq. (\ref{prefactor}))
is fixed  unambiguously by considering the following ratios of amplitudes
in the large-$m_e$ limit
\bea
\lim_{{\mathrm{large~}}m_e}\ \ {{\cal A}_{4 \gamma}^{\rm SM} \over {\cal
A}_{P}^{\rm SM}}=
{{\cal A}_{4 \gamma}^{\rm eff} \over {\cal A}_{P}^{\rm eff}}\,,
\eea
where $P$ stands for our 5-leg processes.

\subsection{Cross sections}

Using the obtained effective Lagrangian eq. (\ref{our}) that coincides with
the one used in \cite{dic3}
we have evaluated the cross section for the first two processes:
\bea
\sigma^{{\mathrm{eff}}}(\gamma\nu\to\gamma\gamma\nu)=
{262 \over 127575}{G_F^2 a^2 \alpha^3\over  \pi^4 }
\left({\omega\over m_e}\right)^8 \omega^2\label{tot}\,. \nonumber
\eea
\bea
\sigma^{{\mathrm{eff}}}(\gamma\gamma\to \gamma\nu\bar\nu)={2144  \over 637875}
{G_F^2 a^2 \alpha^3\over  \pi^4 }
\left({\omega\over m_e}\right)^8 \omega^2\,.
\label{resint}
\eea
In \cite{nous} we have also computed the unpolarized differential cross
sections and the polarized cross sections, using as a check for each polarized
cross section of the second process the well-known polarized cross sections
of the 4-photon scattering process.
The results in \cite{nous} coincide for the differential and total cross 
sections with the ones obtained in \cite{dic3} and are, therefore, 
in disagreement with
the ones obtained by Shabalin et al. in \cite{Shabalin}.

\section{Direct computation of the 5-leg processes}

The range of energy in which the  processes (\ref{eq4}) are relevant
is well below the $W$ mass, so that we treated the neutrino--electron 
coupling as a four-Fermi interaction. We also assumed massless neutrinos.
The total amplitude reads
\bea \label{totamp}
\!\!M(\lambda, \rho, \sigma)\!=\! -2 (1+v_{e})\, \Gamma'_{\mu}
     \sum_{i=1}^3 I_{i}^{ \mu}(p_1, p_2, p_3, \lambda,\rho, \sigma)
     \eea
 $I_{2}^\mu$ and $I_{3}^\mu$ are obtained from $I_{1}^{\mu}$ by the
replacements $p_{2} \leftrightarrow
p_{3}$ and $ \epsilon(p_2,\rho) \leftrightarrow 
\epsilon(p_3,\sigma) $ for $I_{2}^\mu$, and 
$p_{2} \leftrightarrow p_{1}$ and 
 $\epsilon(p_2,\rho) \leftrightarrow \epsilon(p_1,\lambda)$ for $I_{3}^\mu$. 
 The factor $\Gamma'_\mu$ is  
\bea
\Gamma'_\mu \!= \!{(g s_{W})}^3 {\left( \!{g \over 2 c_{W}} \right)\!}^2 \!
\left({1\over \Delta_{Z}} \right)\!
{1 \over {(2 \pi)}^{4}}\,{\bar v}_{+}(5) \gamma_{\mu}u_{-}(4).
\eea 
A detailed description of the method used for the computation of the total
amplitude eq. (\ref{totamp}) can be  found in \cite{nous2}. 
The reduction of $M(\lambda, \rho, \sigma)$ to scalar one-loop
integrals is  performed with the help 
of the technique described in \cite{Pittau}. 
The general philosophy of such a method is using the $\gamma$ algebra 
in the traces to reconstruct the denominators appearing in the loop integrals,
rather than making a more standard tensorial decomposition \cite{Passarino}.
The algorithm can be iterated in such a way that only scalar and
rank-one functions appear at the end of the reduction, at worst
together with higher-rank two-point tensors. In that way, 
all the results are expressed only in terms of
scalar functions with 3 and 4 denominators, 
 rank-1 integrals with 3 and 4 denominators,
 rank-2 integrals with 3 denominators,   and 
 rank-3 functions with 3 denominators. 
This already provides an important simplification with respect
to the standard decomposition, in that the computation of tensors such as
\bea
T^{\mu \nu;\, \mu \nu \rho;\, \mu \nu \rho \sigma} =
\int d^nq \,{q^\mu q^\nu ;\, q^\mu q^\nu q^\rho;\, q^\mu q^\nu q^\rho q^\sigma  \over
D_0\,D_{-1}\,D_2\,D_{(23)}}
\eea 
is completely avoided. A suitable choice of the
polarization vectors \cite{kleiss} as explained in \cite{nous2} is 
the key ingredient in the case at hand. To obtain compact expressions, 
we made a large use of the Kahane--Chisholm manipulations over 
$\gamma$ matrices \cite{kahane}. Such identities
are strictly four-dimensional, while we are, at the same time,  
using dimensional regularization.
Our solution is splitting, {\em before any trace manipulation}, 
the $n$-dimensional integration momentum appearing in the traces 
as \cite{Pittau} $q\to q +{\tilde{q}}$, 
where $q$ and ${\tilde{q}}$ are the four-dimensional and 
$\epsilon$-dimensional components ($\epsilon= n-4$), respectively, 
so that $q \cdot \tilde{q}= 0$.
The $\gamma$ algebra can then be safely performed in four dimensions, 
at the price of having additional terms.
In fact, the splitting $q\to q +{\tilde{q}}$ is equivalent to
 redefining $m_e^2\to m_e^2- {\tilde{q}}^2$ from
the beginning. The net effects are then the  extra integrals
 containing powers of $\tilde{q}^2$ in the numerator, 
whenever $m_e^2$ is present in the formulae.
The computation of such integrals in the limit $\epsilon \to 0 $ is 
straightforward \cite{Pittau}.  Finally,
a standard Passarino--Veltman decomposition \cite{Passarino}
of the simple remaining tensorial structures in terms of scalar 
loop functions, concludes our calculation. We implemented the
outcoming formulae in a Fortran code, performing the phase-space
integration by Monte Carlo. Numerical results are reported 
in the next section.

\subsection{Results}
Our formulae remain valid also when including all neutrino species.
In this case, only the first diagram in {Fig. 2a} contributes, at 
leading order in ${\omega}/{m_e}$,
because the second one is suppressed by powers of $\omega/m_{\mu,\tau}$. 
Therefore, the
inclusion of all neutrinos can be achieved by simply replacing $(1+v_e)$ 
with $(1+ 3\,v_e)$ in \mbox{eq. (\ref{totamp})}.
However, we only considered $\nu_e$ in our numerical results.
Effective and exact computations are compared in 
Figs. 3, 4 and 5, 
 for the three processes.
Furthermore, Table 1 shows the ratio between exact cross section 
and $\sigma^{\mathrm{eff}}$ for several
values of $\omega/m_e$, then exhibiting   the range of validity
 of the effective theory.  

From the above figures and numbers it is clear that, for all three processes, 
the effective theory is valid only when, roughly, $\omega/m_e \leq 2$, 
as expected. At larger $\omega$, the exact
computation predicts a softer energy dependence with respect to the 
$\left({\omega}/{m_e}\right)^{10}$ behaviour given by the effective
Lagrangian. 

All these results are in full agreement with those reported 
in \cite{dic1}.
In order that the results be useful, the only \footnote{A large-$m_e$
expansion
 has been performed. The first term does not give numerical predictions  that are
 useful to extend the effective theory beyond $\omega=m_e$.} option is 
to fit the curves in Figs.
3, 4 and 5.
The results of the fits are
\bea\label{fit1}
&&{\sigma(\gamma\nu\to\gamma\gamma\nu)\over
\sigma^{{\mathrm{eff}}}(\gamma\nu\to\gamma\gamma\nu)}=
 r^{-2.76046}~\times
{\rm exp}[2.13317-\nn \\ &&\!\!2.12629 {\rm log}^{2}(r) +0.406718 {\rm log}^{3}(r)  
-0.029852{\rm log}^{4}(r)],  \nonumber \\
&&{\sigma(\gamma\gamma \to \gamma\nu {\bar \nu})\over
\sigma^{{\mathrm{eff}}}(\gamma\gamma \to \gamma\nu {\bar \nu})}=
r^{-7.85491}~\times
{\rm exp}[4.42122+\nn\\&&\!\!0.343516 {\rm log}^{2}(r)-0.114058 
 {\rm log}^{3}(r)  
+0.0103219{\rm log}^{4}(r)], \nonumber \\
&&{\sigma(\nu {\bar \nu}\to \gamma \gamma \gamma)\over 
\sigma^{{\mathrm{eff}}}(\nu {\bar \nu}\to \gamma \gamma \gamma)}=
 r^{-6.57374}~\times
{\rm exp}[5.27548-\nn\\&&\!\!0.689808 {\rm log}^{2}(r)+0.15014  
{\rm  log}^{3}(r) 
-0.0123385  {\rm log}^{4}(r)]\nonumber
\\ 
\eea
where the effective cross sections $\sigma^{{\mathrm{eff}}}$ are given in 
eq. (\ref{resint}) and ref. \cite{dic3} and $r= \omega/m_{e}$.
All the above fits are valid in the energy range $1.7 < r < 100 $.
\begin{table}\begin{tabular}{|c|c|c|c|}
\hline
$\omega/m_e$&$\gamma\nu\to\gamma\gamma\nu$&$\gamma\gamma\to\gamma\nu\bar\nu$&$\nu\bar\nu\to\gamma\gamma\gamma$ \\ \hline 
0.3 &0.969(8)  &1.09(1)  &1.20(1)  \\ \hline
0.4 &0.923(6)  &1.17(1)  &1.37(1)  \\ \hline
0.5 &0.888(6)  & 1.28(1) &1.68(1)  \\ \hline
0.6 &0.852(4)  &1.47(1)  &2.20(1)  \\ \hline
0.7 &0.826(5)  &1.80(1)  &3.17(1)  \\ \hline
0.8 &0.811(5)  &2.41(1)  &5.31(2)  \\ \hline
0.9 &0.819(6)  & 3.95(2) &11.88(3) \\ \hline
1.0 & 0.880(7) &23.1(1)  &176.3(3) \\ \hline
1.1 &1.19(1)   &18.2(1)  &94.7(2)  \\ \hline
1.3 &1.71(2)   &8.31(5)  &31.6(1)  \\ \hline
1.5 &1.44(1)   &3.37(2)  &11.9(1)  \\ \hline
1.7 &0.996(8)  &1.40(1)  &4.96(3)  \\ \hline
1.9 &0.635(4)  &0.622(3) &2.23(1)  \\ \hline
2.0 &0.503(3)  &0.424(2) & 1.54(1) \\ \hline
\end{tabular}
\vspace{0.5cm}
\hspace {-0.3cm}\caption{{Ratio between exact and effective results for 
the three cross sections. 
The error on the last digit comes from the phase-space Monte Carlo 
integration.}}\label{tab-exact}
\end{table}

\vskip -0.1cm
\vspace{-0.2cm}
\noindent\mbox{\epsfig{file=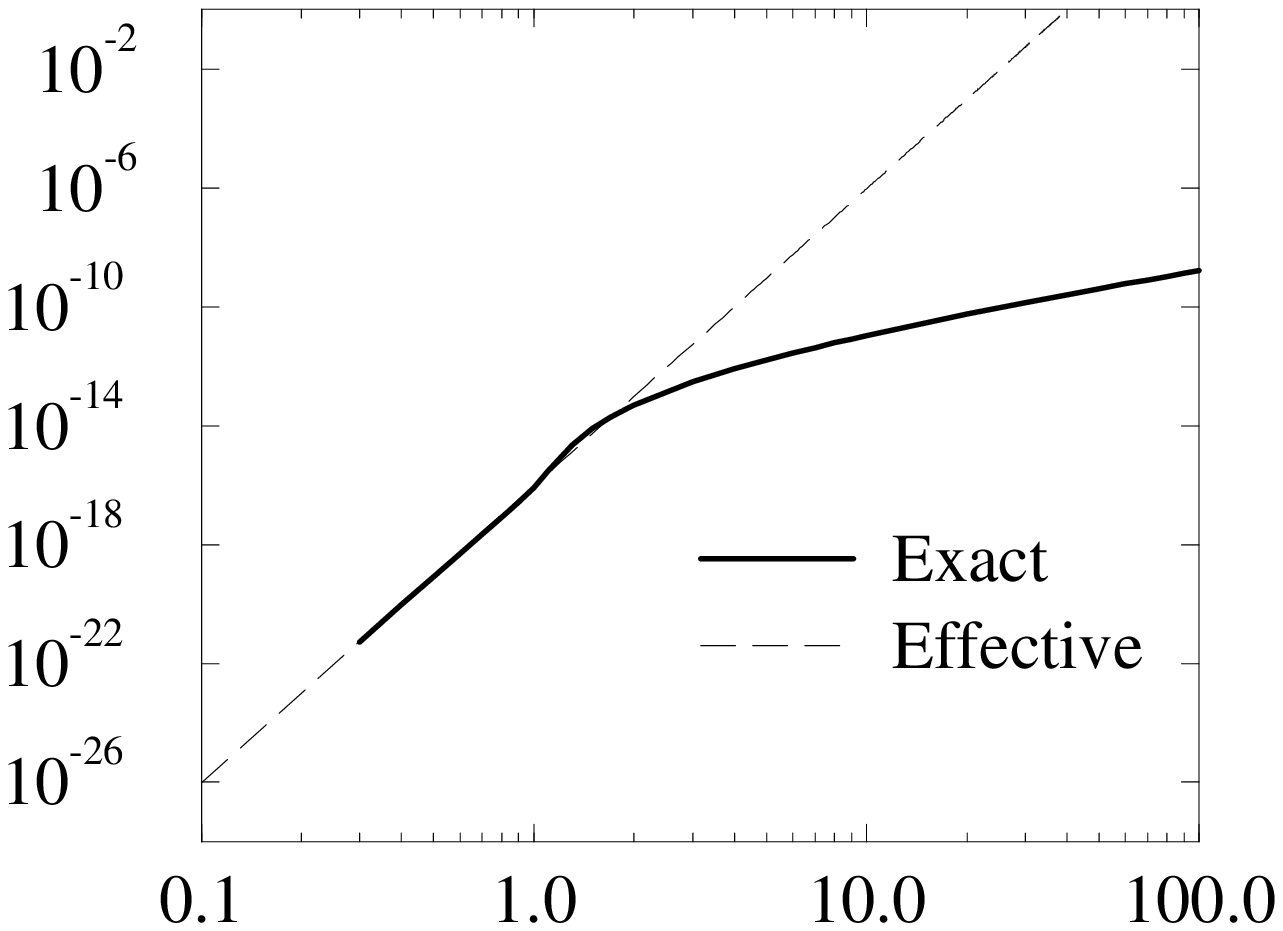,width=3.5in,height=3in}}
\vskip -0.2cm
\vspace{-1cm}
\noindent{{{\small{\bf{Figure 3}}}: 
{\small{ \hskip -0.1cm$\gamma\nu\to\gamma\gamma\nu$ cross section in fb 
as a function of $\omega/m_e$.}}} \protect\label{complet1}
\vspace{-0.7cm}

\noindent\mbox{\epsfig{file=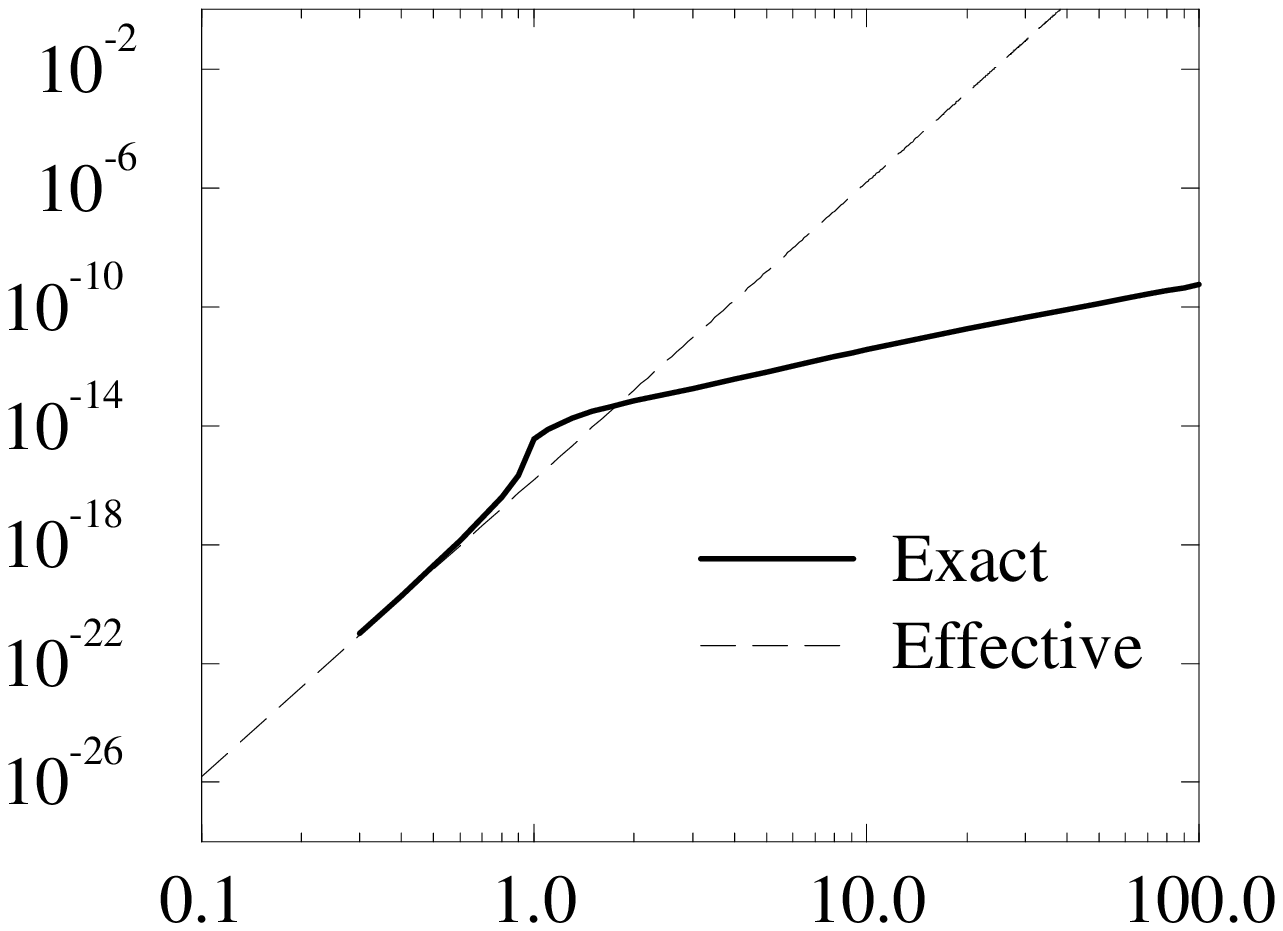,width=3.5in,height=3in}}
\vskip -0.2cm
\vspace{-1cm}
\noindent{{{\small{\bf{Figure 4}}}: 
{\small{  \hskip -0.1cm$\gamma\gamma\to\gamma\nu\bar\nu$ cross section in fb 
as a function of $\omega/m_e$.}}} \protect\label{complet3}
\vspace{-0.5cm}
\noindent\mbox{\epsfig{file=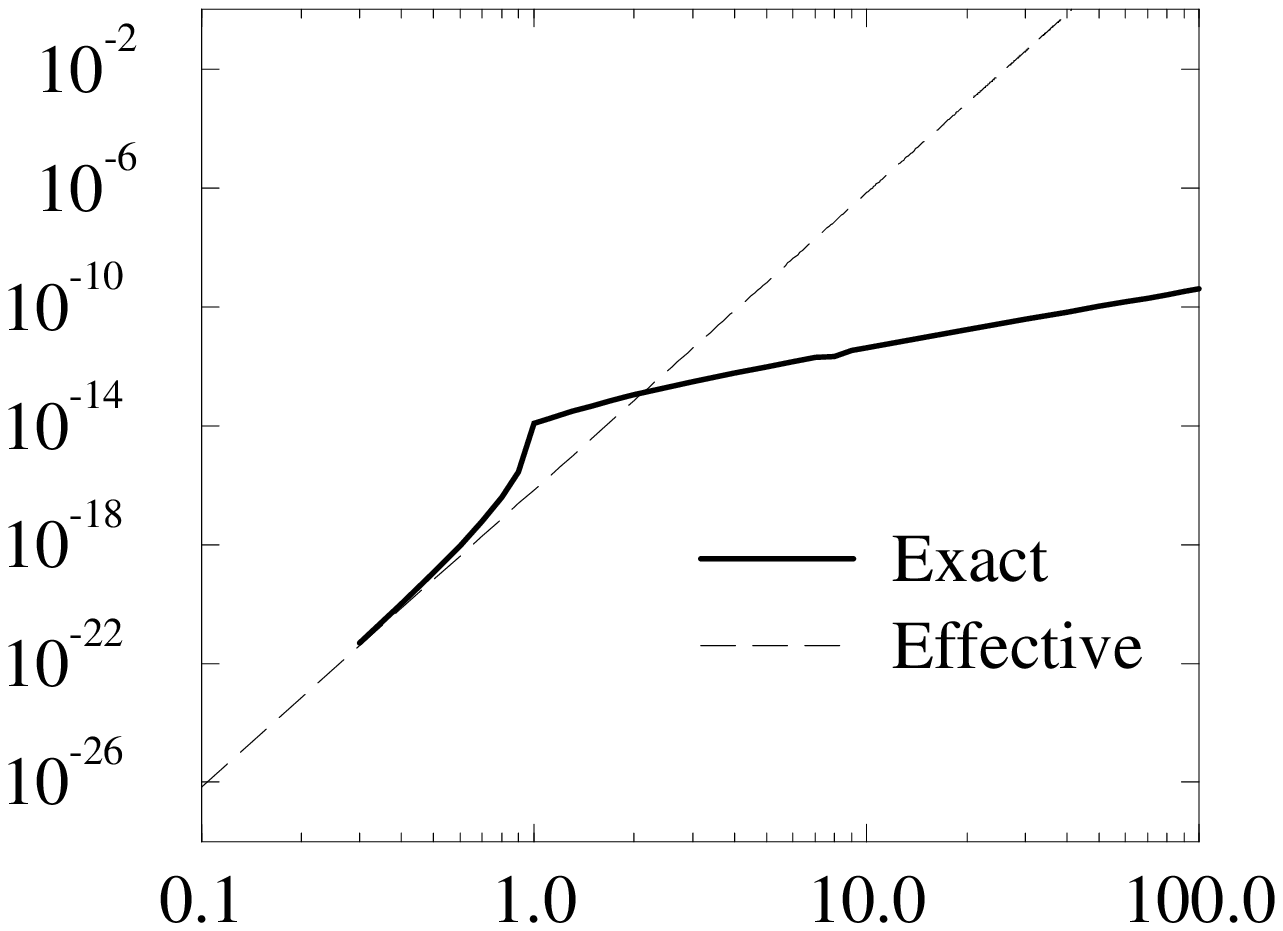,width=3.5in,height=3in}}
\vskip -0.1cm\vspace{-1cm}
\noindent{{{\small{\bf{Figure 5}}}: 
{\small{  \hskip -0.1cm$\nu\bar\nu\to\gamma\gamma\gamma$ cross section in fb 
as a function of $\omega/m_e$.}}} \protect\label{complet3}

\section{Astrophysical and cosmological  possible \\implications}\label{conclusion}
The 5-leg  photon--neutrino processes are of interest in astrophysics. The 
processes 
$\nu\gamma\to\nu\gamma\gamma$ and  
$\nu\bar\nu\to\gamma\gamma\gamma$  can affect the mean free path of
 neutrinos
inside the supernova core, while the process  $\gamma\gamma\to\gamma\nu\bar\nu$
 is a possible cooling mechanism for hot objects \cite{teplitz}.

Basing their results on the assumption
\bea\label{steplitz}
    \sigma(\gamma\nu\to\gamma\gamma\nu)= 
    \sigma_0 \left(\frac{\omega}{1~{\rm MeV}} \right)^\gamma\,,
~\sigma_0= 10^{-52}\,{\rm cm}^2\,,
\eea
and on the data collected from supernova 1987 A, the authors of  \cite{teplitz} 
fitted the exponent $\gamma$ in 
\mbox{eq. (\ref{steplitz})} to be less than $8.4$, for $\omega$ of the order
of a few MeV. The physical requirement behind this is that neutrinos 
of a few MeV should immediately leave the supernova, 
so that their mean free path is constrained to be larger than $10^{11}$ cm.

The effective theory predicts $\gamma= 10$, while, using the curves from the
exact calculation, a softer energy dependence is
observed in the region of interest (see Fig. 3). 
A fit to the exact curve 
gives $\gamma \sim 3$ for $1~{\rm MeV} < \omega < 10~{\rm MeV}$,
thus confirming the expectations of \cite{teplitz}.

A second interesting quantity is the range of parameters for which
 the neutrino mean free path for such reactions
is inside the supernova core (of 10 km typical size),
 therefore affecting its dynamics.
Always in \cite{teplitz} it was found, with 
the help of Monte Carlo simulations, that for several choices 
of temperature and
chemical potential, and assuming the validity of the effective theory
 ($\gamma= 10$), this happens when
$\omega \ge 5$ MeV. Since the exact results are now available,
 it would be of extreme interest 
to see how the above prediction is affected. 
More in general, we think that the reactions
in  (\ref{eq4}) should be included in 
supernova codes.

On the basis of the effective theory results, the authors of  \cite{dic3},
 suggested that these processes  
could also have some relevance in cosmology.  
Consider, in fact, the mean number $\bar N$ of neutrino collisions,
via the first of processes  (\ref{eq4}),
in an expansion time $t$ equal to the age of the Universe \cite{Peebles}: 
\bea
\bar N=\sigma(\gamma\nu\to\gamma\gamma\nu) n_\nu\  c \ t\,,
n_\nu={\rm neutrino~number~density}.\nonumber
\eea 
By writing  $n_\nu$ and $t$ in terms of the photon energy at thermal 
equilibrium ($\omega \sim k T$), expressed in units of \\
$10^{10}$ K and denoted 
by $T_{10}$, we get
\bea
n_\nu=1.6 \times 10^{31} T_{10}^3\ {\rm cm}^{-3}\,, 
~~t=2 T_{10}^{-2}\,\, {\rm s}.
\eea
When $\bar N$ is large, the neutrinos are in thermal contact with matter and radiation, while, for 
$\bar N\sim 1$ (namely $\sigma\sim 10^{-42}
T_{10}^{-1}\,\,\mathrm{cm}^{2}$), 
the neutrinos decouple. 
Applying the effective formula in \mbox{eq. (\ref{resint})},
 the resulting decoupling 
temperature
is $T_{10}\ \sim\ 9.5$, namely $\omega\ \sim\ 8.2$ MeV, therefore outside the range of validity
of the effective theory. By repeating the same analysis with the exact result, we found instead
that $\bar N$ becomes of the order of 1 at $\omega\ \sim\ 1$ GeV, and 
at these energies, other processes enter the game. 
In conclusion, the five-leg reactions 
in \mbox{eq. (\ref{eq4})} are unlikely to be important for a study of the neutrino
decoupling temperature, contrary 
to what the effective theory seemed to suggest.
\section*{Acknowledgements}
We thank the authors of \cite{dic1} for having informed us about their recent
computation. \\We also wish to thank M.B. Gavela, G.F. Giudice and O. P\`ene
for helpful remarks.

\end{document}